\documentclass[%
 reprint,
superscriptaddress,
 amsmath,amssymb,
 aps,
twocolumn
]{revtex4-1}

\usepackage{graphicx}
\usepackage{dcolumn}
\usepackage{bm}

\begin{document}

\preprint{APS/123-QED}

\title{Superspreading k-cores at the center of COVID-19 pandemic persistence}

\author{Matteo Serafino}
\affiliation{
    Levich Institute and Physics Department, City College  of New York, New York, NY 10031, USA
}
\affiliation{
    IMT School for Advanced Studies, 55100 Lucca, Italy
}
\author{Higor S. Monteiro}
\affiliation{
    Departamento de F\'isica, Universidade Federal do Cear\'a, 60451-970 Fortaleza, Cear\'a, Brazil
}
\author{Shaojun Luo}
\affiliation{
    Levich Institute and Physics Department, City College  of New York, New York, NY 10031, USA
}
\author{Saulo D. S. Reis}
\affiliation{
    Departamento de F\'isica, Universidade Federal do Cear\'a, 60451-970 Fortaleza, Cear\'a, Brazil
}
\author{Carles Igual}
\affiliation{
    Instituto de Telecomunicaciones y Aplicaciones Multimedia (ITEAM),
    Departamento de Comunicaciones, Universitat Polit\`ecnica de Val\`encia, Val\`encia 46022, Spain
}
\author{Antonio S. Lima Neto}
\affiliation{
    Department of Epidemiological Surveillance, Fortaleza Health Secretariat, Fortaleza, Cear\'a, Brazil
}
\affiliation{
    Department of Public Health, University of Fortaleza Medical School, Fortaleza, Cear\'a, Brazil
}
\author{Matias Travizano}
\affiliation{
    Grandata, Inc, 550 15th St. Suite 36, San Francisco, CA 94103, USA
}
\author{Jos\'e S. Andrade, Jr.}
\affiliation{
    Departamento de F\'isica, Universidade Federal do Cear\'a, 60451-970 Fortaleza, Cear\'a, Brazil
}
\author{Hern\'an A. Makse}
\affiliation{
    Levich Institute and Physics Department, City College  of New York, New York, NY 10031, USA
}


\date{}

\begin{abstract}
    The spread of COVID-19 caused by the recently discovered
  SARS-CoV-2 virus has become a worldwide problem with devastating
  consequences. To slow down the spread of the
  pandemic, mass quarantines have been implemented globally, provoking
  further social and economic disruptions. Here, we
  implement a comprehensive contact tracing network analysis to find
  an  optimized quarantine protocol to dismantle the
  chain of transmission of coronavirus with minimal disruptions to
  society. We track billions of
  anonymized GPS human mobility datapoints from a compilation of
  hundreds of mobile apps deployed in Latin America to
  monitor the evolution of the contact network of disease transmission
  before and after the confinements. As a consequence of the
  lockdowns, people's mobility across the region decreases by
  $\sim$53\%, which results in a drastic disintegration of the
  transmission network by $\sim$90\%. However, this disintegration did
  not halt the spreading of the disease. Our analysis indicates that
  superspreading k-core structures persist in the transmission network
  to prolong the pandemic.  Once
  the k-cores are identified,  an optimized strategy to
  break the chain of transmission is to quarantine a minimal number of
  'weak links' with high betweenness centrality connecting the large k-cores.  As
  countries built contact tracing apps to fight the pandemic, our results could
  turn into a valuable resource to help deploy quarantine protocols
  with minimized disruptions.
\end{abstract}

\maketitle

\section{Introduction}

In the absence of vaccine or treatment for COVID-19, state-sponsored
lockdowns have been implemented worldwide to halt the spread of the
ongoing pandemic creating large social and economic disruptions
\cite{who,senpei,vespignani}. In addition, some countries have also
implemented digital contact tracing protocols to track the contacts of
infected people and reinforce quarantines by targeting those at high
risk of becoming infected
\cite{mit,DeVries,Singapore,Israel,Helsenorge,Noem,mit2,iceland,Ferretti_2020_Quantifying,sciadv}.  Here we develop,
calibrate, and deploy a contact tracing algorithm to track the chain
of disease transmission across society. We then search for
quarantine protocols to halt the epidemic spreading with minimal
social disruptions
\cite{epidemic,vespignani-pre,newman,barabasi-attack,shlomo,ci}.

Our study uses two complementary datasets. The first includes data
from 'Grandata-United Nations Development Programme partnership to
combat COVID-19 with data' \cite{grandata}. It is composed of
anonymized global positioning system (GPS) data from a compilation of
hundreds of mobile applications (apps) across Latin America that allow
to track the trajectories of people (users).  The data identify each
mobile phone device with a unique encrypted mobile ID and specifies
its latitude and longitude location through time, encoded by geohash
with 12 digits precision. Typically, this dataset generates $\sim$ 450
million data points of GPS location per day across Latin America in
particular in the state of Cear\'a, Brazil (see SM sections~I-V).

The second dataset is an anonymized list of confirmed COVID-19
patients obtained from the Health Department authorities from both
states. It includes the geohash of the address, the SARS-COV-2 test
detection date and first day of symptoms of COVID-19.
We cross-match the geolocation of the patients with the GPS dataset
obtaining the encrypted mobile ID of the patients (see SM sections~I-V). We
then trace the geolocalized trajectories of COVID-19 patients during a
period -14/+7 days from the onset of symptoms to look for contacts of
the infected person to define the transmission network using the model
described below.

\section{COVID-19 model}

\begin{figure}[!t]
 \includegraphics[width=\columnwidth]{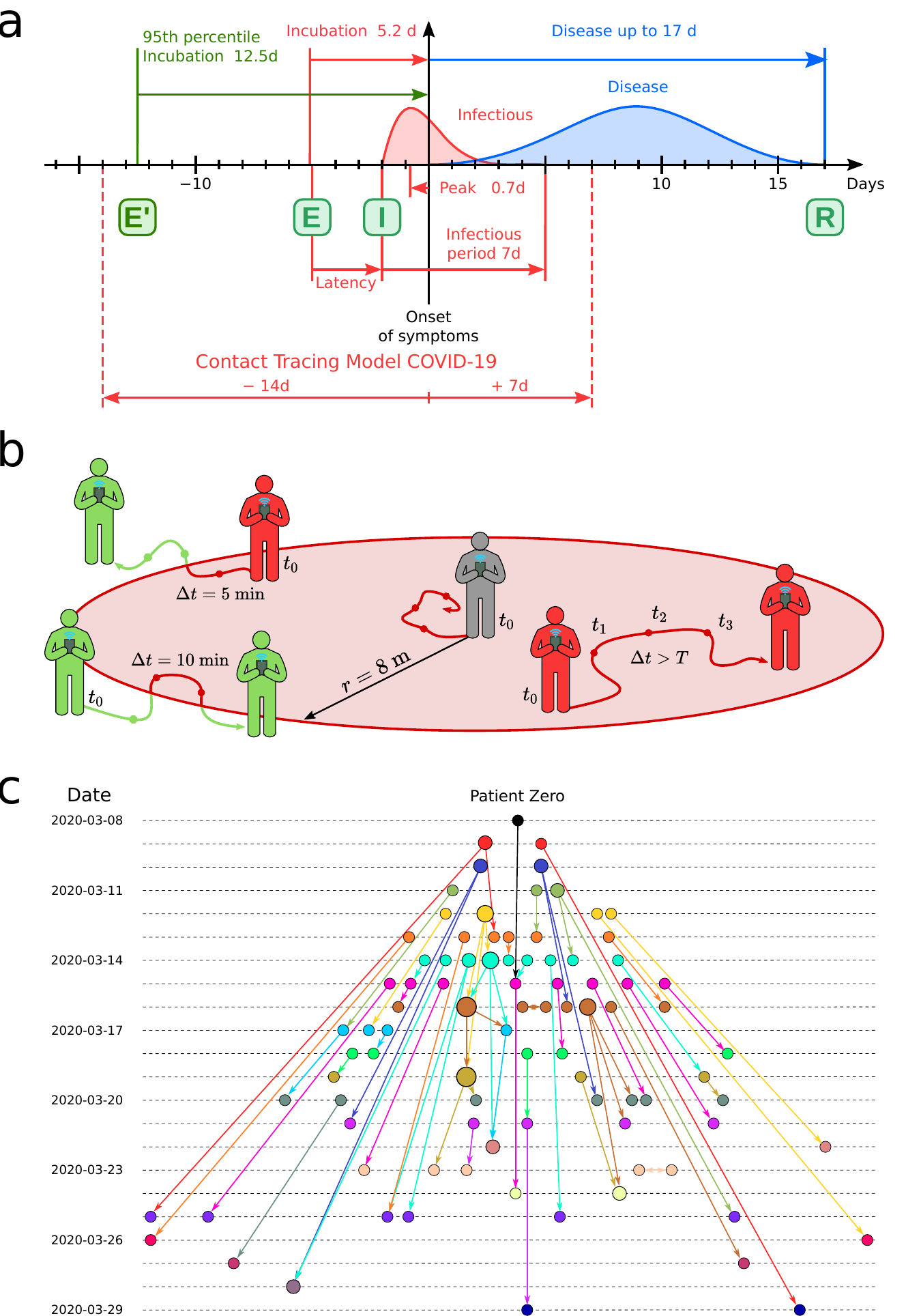}
  \caption{{\bf COVID-19 contact model.} {\bf (a)}
Infectiousness profile of COVID-19. The COVID-19 pandemic is
represented with a SEIR model. From exposure (E) the virus is
incubated in average for 5.2 days (12.5 days $95^{th}$ percentile),
starting the symptoms 2 days after infectiousness (I) and lasting the
disease up to 17 days to recover (R). We use a window -14/+7 days from
the first symptoms to detect infectious and exposure.  {\bf (b)}
Contact area used in the contact tracing model. The grey person is at
the first datapoint of the source at $t_0$. We collect all datapoints
for every user in a $T$=30 min forward window ($t_1, t_2, t_3, ...,
t_0+T$) within an 8 m circle from the initial position. For each
target (green and red) we compute the average position and the time
spent inside the contact area (red part of the trajectory line). {\bf
  (c)} Partial transmission tree of outbreak of confirmed SARS-CoV-2
infection identified by contact tracing during calibration in the
month of March 2020. Links goes from the source of infection to the
target. The colors represent the day of first symptoms for each node
and size is the out-degree.}
\label{model}
\end{figure}

The COVID-19 spreading model is represented
by a Susceptible-Exposed-Infectious-Recovered (SEIR) process
\cite{epidemic}
(Fig. \ref{model}a).
The infectiousness period of an infected person starts 2 days before
and lasts up to 5 days after the onset of symptoms
\cite{He_2020_Temporal}. In this paper, we add two days to each of
these limits to conservatively capture most transmissions. Thus, in
principle, to trace those people potentially infected by COVID-19
patients, we track contacts 4 days before and 7 days after the
reported date of first symptoms (see Fig. \ref{model}a). In addition,
we extend the tracing period further back in time to also consider
exposures that could come from asymptomatic cases. Exposures start the
incubation period of the infected person which can occur up to 12.5
days before onset of symptoms (5.2 days on average, 95\% percentile
12.5 days \cite{li_nejm,Sanche_2020_High}, Fig. \ref{model}a). To
conservatively trace these exposure events, we add $\sim$2 days to
this incubation period and obtain the widely used 14 days period.
Hence, to trace transmission and exposure cases, we perform contact
tracing over -14/+7 days from onset of symptoms
(Fig. \ref{model}a). We note that the peak of infectiousness as well
as 44\% (95\% confidence interval, 25-69\%) of infected cases occur
during the pre-symptomatic stage \cite{He_2020_Temporal}.  Thus,
performing contact tracing is essential to stop the spreading of the
disease.

\section{Contact model}

The GPS geolocation of the trajectories of
both infected and susceptible people is used to trace several layers
of contacts in the transmission network using the following model. A
contact at time stamp $n$ is initiated with an infected user (source)
at time $t_0$ (see Fig. \ref{model}b). At $t_0$ we draw a contact area
as a circle centered in the source position with a radius $r$. We then
gather all the GPS datapoints from susceptible users (targets) that
enter the contact area from $t_0$ to $t_0 + T$, where $T$ is the total
exposure time. We follow the trajectories of source and target within
the time-space area and compute the probability of infection at time
stamp $n$ as $p_i[n] = p_d[n] \cdot p_t[n]$, where $p_d[n]$ is the
spatial component, and $p_t[n]$ is the temporal component. When the
average overlap between source and target is zero, then $p_d[n]=1$,
and when the overlap is 2$r$, then $p_d[n]=0$. On the other hand, when
the exposure time $\ge T$, then $p_t[n]=1$, and decreases to
$p_t[n]=0$ as the exposure time decreases (see SM sections~I-V for
definitions).   The probability $p_d[n]$ quantifies the
  contact probability for two users in the same area defined by $r$.
A contact requires non only a space overlapping but also a time
overlap, $p_t[n]$, which quantifies the probability that two users met
based on the time commonly spent in the same area. We then combine
these two probabilities for each timestamp $n$ into their product.



Contacts with low probability of infection $p_i[n]$, but repeated
throughout time, can also infect the target. To incorporate this
effect in the model, we define the probability of infection for a
series of repeated contacts $P_i[n]$ as a recursive formula from time
1 to $n$ with $P_i[0]=0$:
\begin{equation}\label{recursive}
P_i[n] = p_i[n](1-P_i[n-1]) + P_i[n-1].
\end{equation}
The iteration of contacts between source and target, $P_i[n]$,
generates higher probability of infection than a single contact
$p_i[n]$. This means that there is a difference between a short single
contact between two people and short repeated contacts between the
same people. The latter scenario should have a larger probability than
the former to become infected.  While the distribution of $p_i[n]$ is
homogenous without a clear threshold for an infectious contact,
$P_i[n]$ presents a very polarized distribution where the values are
accumulated in the extremes: $P_i$ = 0 or $P_i$ = 1 (see SI
Fig.~SIA).  Thus, $P_i[n]$ is better indicator
than $p_i[n]$ to separate infectious from non-infectious contacts. A
contact is then considered infectious when this probability exceeds a
certain threshold, $P_i[n] > p_c$. The hyperparameters of the contact
model $(T, r, p_c)$ are obtained by calibrating the model using only
the contacts between infected people to reproduce the basic
reproduction number $R_0=2.78$ in Cear\'a in the month of March, 2020
(see SM sections~I-V). We obtain $T$ = 30 min, $r$ = 8 m and $p_c$ =
0.9. Thus, a contact is defined with probability one when exposure is
at least 30 minutes within a distance $\ll 8$m. This calibration
procedure provides the partial transmission tree of the outbreak from
patient zero to the end of the calibration period shown in
Fig. \ref{model}c.

\section{Transmission network model}

Next, we create the contact
network of coronavirus transmission by first tracing the trajectories
of confirmed COVID-19 patients to search for contacts -14/+7 days from
the onset of symptoms using the above model. From the first contact
layer, we add four layers of contacts to constitute the contact
network of transmission that is used to monitor the progression of the
pandemic.  The time-varying network is aggregated to a snapshot
defined over a time window of a week \cite{epidemic} (SM Section~S6.1). 
We find that other aggregation windows give
similar results as presented.

Next, we analyze the spatio-temporal properties of the contact
network. The government of the State of Cear\'a imposed a mass
quarantine on March 19, 2020 which led to a decrease in people's
mobility by 56.5\% as shown in Fig. \ref{gcc}a. During the lockdown,
only the displacements of essential workers were allowed.
A large decrease in mobility is also observed across all Latin
America, see \cite{grandata}.

\section{Giant connected component (GCC)} 

\begin{figure}[!t]
  \centering
   \includegraphics[width=0.9\columnwidth]{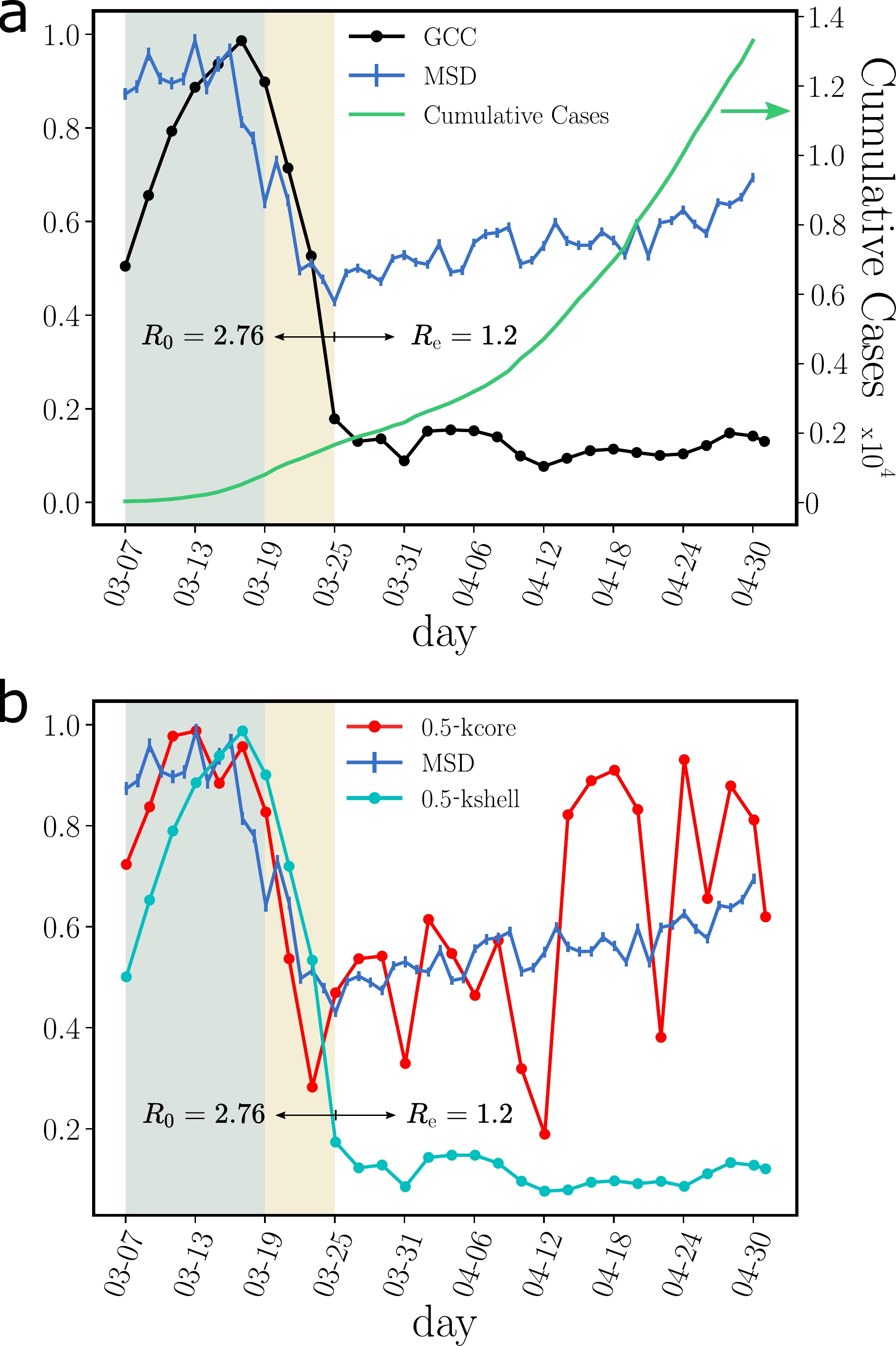}
   \caption{{\bf Structural components of transmission networks
  across the lockdown.} {\bf (a)} Evolution for different metrics in
Cear\'a, Brazil, previous to the mass quarantine (grey area), right
after the imposed quarantine (yellow area) and later. The plot shows
the root mean square displacement (MSD) normalized by the maximum
value over the total period (blue), the cumulative number of cases
(green) and the size of the GCC normalized by the maximum value over
the total period (black). The uncertainty corresponds to the standard
error (SE).
The mobility data is showcased in the Grandata-United Nations
Development Programme map shown in {\it https://covid.grandata.com}.
The initial rise in GCC is due to the lack of data before March 1.
{\bf (b)} The plot shows the 0.5-kcore size (red), the 0.5-kshell size
(cyan) all normalized by their respective maximum value
pre-lockdown. While the size of the 0.5-kshell is reduced drastically
during the lockdown, the 0.5-kcore was not reduced as much and keeps
increasing, contributing to sustain the pandemic. The 0.5-kcore seems
to follow the trend in the MSD, which we plot again to show this
trend.}
   \label{gcc}
\end{figure}

To understand the effect of
the lockdown on the contact network, we think by analogy with a 'bond
percolation' process \cite{caldarelli,epidemic,newman}.
In bond percolation, the network connectivity is reduced by removing a
small fraction of links (bonds) between nodes, and the global
disruption in network connectivity is monitored by studying the
normalized size of the giant connected component (see
Methods). Following this analogy, the lockdown acts as a percolation
process, and therefore we monitor the GCC of the transmission network
before and after the lockdown. We find a  large decrease
  in the size of the GCC
\cite{caldarelli,epidemic,epidemic} within 6 days of the
implementation of the lockdown on March 19, when the GCC is almost
fully dismantled decreasing by 89.6\% of its pre-lockdown size
(Fig. \ref{gcc}a).

Despite the disintegration of the GCC, the cumulative number of cases
kept growing albeit at a lower rate (Fig. \ref{gcc}a).
We find that the mass quarantine was able to reduce the basic
reproduction number from $R_0=2.78$ before lockdown to an effective
reproduction number of $R_e=1.2$ after the lockdown (Fig. \ref{gcc}a).
Despite this disruption in the network connectivity, $R_e$ has not
decreased below one, as it would have been needed to curb the spread
of the disease.

\begin{figure}[ht]
   \centering
         \includegraphics[width=\linewidth]{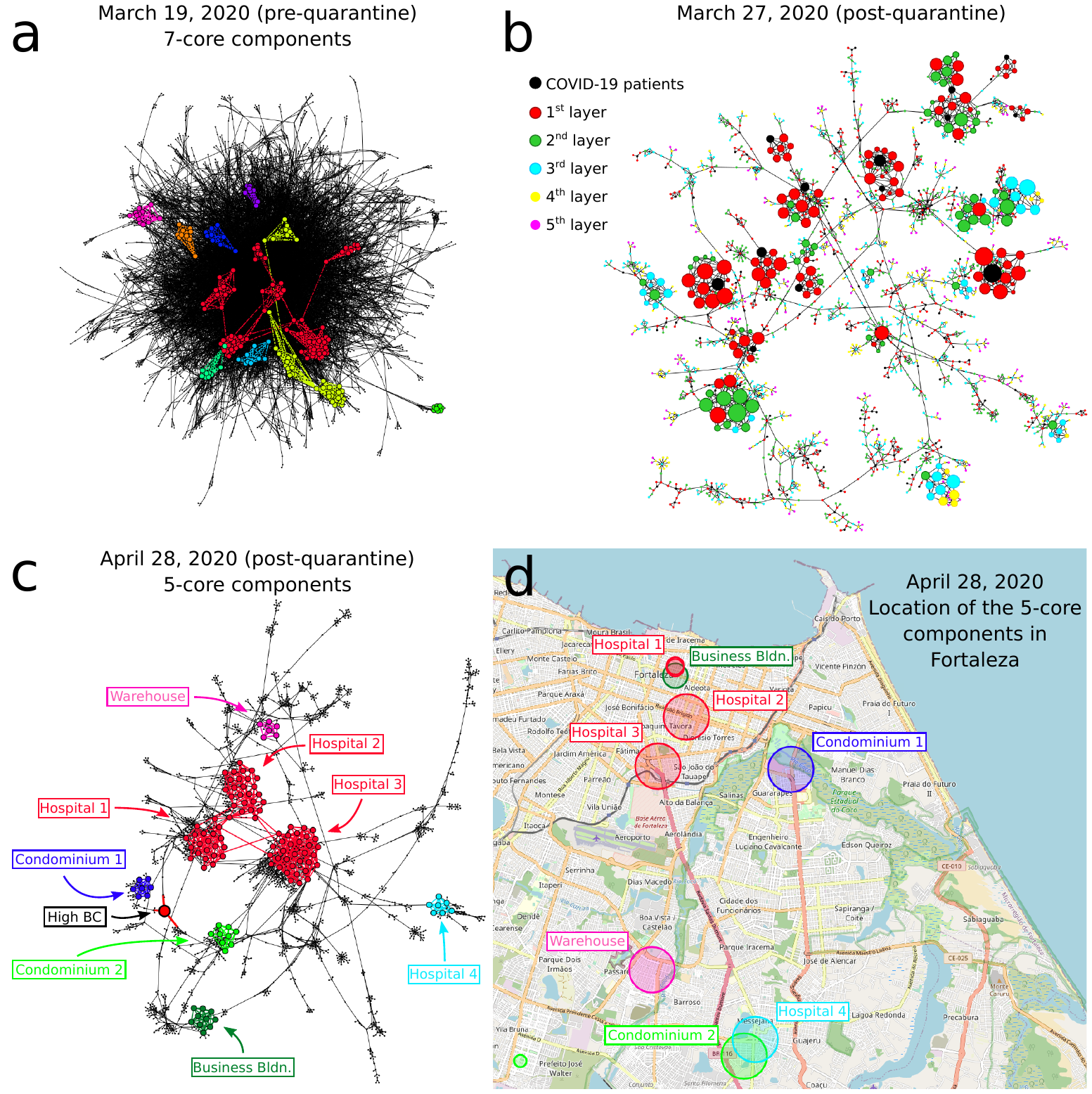}
  \caption{{\bf Evolution of GCC and k-cores over the
  quarantine.} Disease transmission networks in the state of Cear\'a
over time before and after the lockdown on March 19, 2020.  {\bf (a)}
Transmission network on March 19 (pre-lockdown). A hairball
highly-connected network is observed. The disconnected components of
the 7-core ($k_{\rm core}^{\rm max}=12$ in this network) are
colored. These components are well connected into the hairball network
as expected since mobility and connectivity is high.  {\bf (b)} The
pre-quarantine hairball in {\bf (a)} has been untangled and the
k-cores have emerged 8 days into the lockdown on March 27. Here, we
color the nodes according to layers of the transmission network
starting at COVID-19 patient (black nodes).  Size of nodes is
according degree.  {\bf (c)} Network on April 28 including the
components of the 5-core in different colors ($k_{\rm core}^{\rm
  max}=7$ for this network). Visible is the high betweenness
centrality node representing the weak-link of this k-core.  {\bf (d)}
We plot the location of the contacts in the map of Fortaleza
constituting the components of the 5-core of the April 28 in {\bf
  (c)}. The size of the circles in the map corresponds to the number
of contacts inside each location. The colors correspond to the
clusters of the 5-core in {\bf (c)}. The 5-core sustaining
transmission is composed of clusters of contacts localized in
hospitals, large warehouses and business buildings.
Hospital 3, one of the largest in Fortaleza, constitutes the maximal
$k_{\rm core}^{\rm max}=7$ of the pandemic.}
  \label{networks}
\end{figure}

The drastic reduction in the GCC is visually apparent in the contact
networks in Fig. \ref{networks}. Before lockdown on March 19
(Fig. \ref{networks}a), the network is a strongly-connected
unstructured 'hairball'.  Eight days into the lockdown on March 27
(Fig. \ref{networks}b), the network has been untangled into a set of
strongly-connected modules integrated by tenuous paths of
contacts. This structure is even more pronounced a few weeks later on
April 28 (Fig. \ref{networks}c).



\section{Superspreading k-core structures}

    \begin{figure*}[!t]
      \centering
      \includegraphics[width=0.75\linewidth]{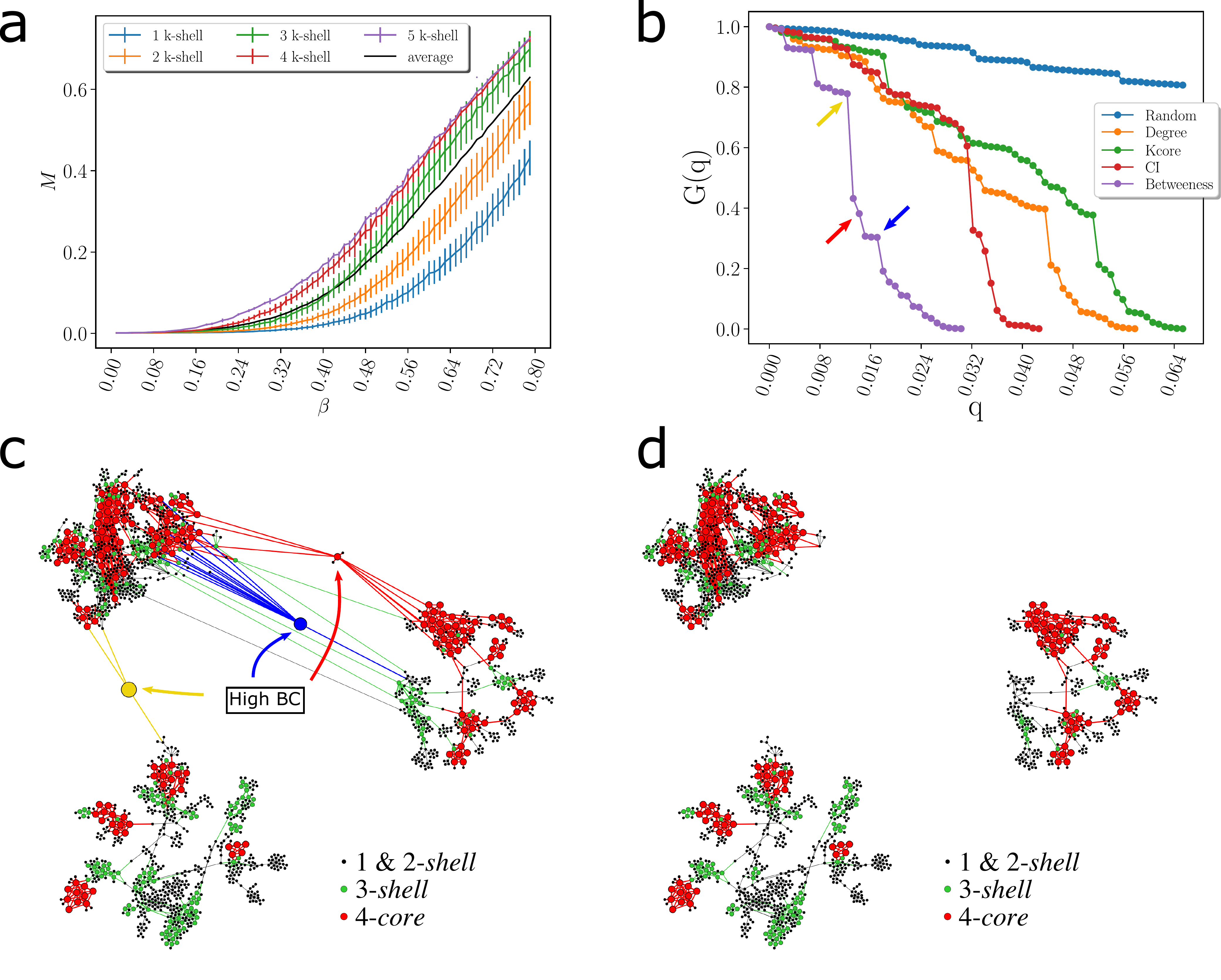}\\
\caption{ {\bf Weak links and k-cores}.  {\bf (a)}
Average size of infected population, $M$
\cite{kcore}, in an outbreak average over all starting nodes in a
k-shell as a function of the probability of infection $\beta$ for a
SIR model on the network in Fig. \ref{networks}c during the
lockdown. The black is the average value over all the network. The
average divides the k-shell contribution to the spreading of the virus
in two groups: above and below the average. The 0.5-kcores have
maximal spreading and the 0.5-kshell have minimal spreading. Error
bars correspond to a confidence interval of 95\%. {\bf
  (b)} Optimal percolation analysis performed over the network in
Fig. \ref{networks}c during the lockdown in following different attack
strategies and their effect on the size of the largest connected
component $G(q)$ versus the removal node fraction, $q$. Nodes are
removed (in order of increasing efficiency): randomly (blue); by the
highest k-shell followed by high degree inside the k-shell
\cite{kcore}; by highest degree (orange); by collective influence
(red) \cite{ci}; and by the highest value of betweenness centrality
(green) \cite{freeman,friedkin91}.  After each removal we re-compute
all metrics. The most optimal strategy among those studied is removing
the nodes by the highest value of betweenness centrality.  {\bf
  (c)-(d)} Effect of removing three high betweenness centrality nodes
shown in Fig. \ref{sir-attack}b in the network of
Fig. \ref{networks}c.  {\bf (c)} We show the 2-core component of the
network after the removal of 12 high betweenness centrality nodes.
The red node is the one with the highest betweenness centrality value
(next node to remove, 13th) and the blue node is the 14th
removal. Different k-cores and k-shell are in different colors.  {\bf
  (d)} Network k-cores are disintegrated after the removal of the high
BD nodes.}
\label{sir-attack}
\end{figure*}

The highly connected
modules found in Fig. \ref{networks}b and \ref{networks}c are k-core
structures \cite{dorogotsev,carmi,alvarez,kcore} of higher complexity
than the GCC (which is a 1-core), that are known to sustain an
outbreak even when the GCC has been disintegrated
\cite{epidemic,kcore}. The k-core of a graph is the maximal subgraph
in which all nodes have a degree (number of connections) larger or
equal than $k$ \cite{dorogotsev,carmi,alvarez,kcore}. The k-shell is
the periphery of the k-core and is composed by all the nodes that
belong to the k-core but not to the (k+1)-core (see SM sections~I-V for
definitions and SM Figs. S2, S3, and
S4).  The k-core is obtained by iteratively
pruning the nodes with degree smaller than $k$. For instance, the
3-core is obtained by removing the 1-shell and 2-shell in a k-shell
decomposition process (see SM Figs. S2,
S3). Thus, all nodes in a k-core have at least degree
$k$, and are connected to other nodes with degree at least $k$
too. K-cores are nested and can be made of disconnected components
(see SM Fig. S4). High k-cores are those with
large $k$ up to a maximal $k_{\rm core}^{\rm max}$, and constitute the
inner most important part of the network. In theory, the high k-cores
are known from network science studies to be the reservoir of disease
transmission persistence \cite{epidemic,kcore}. On the contrary, low
peripheral k-shells (see SM Fig. S2) do not contribute
as much to the spread as the high inner k-cores.

Figure \ref{gcc}b shows that despite the disappearance of the GCC,
there is a significant maximal k-core that was not dismantled by the
mass quarantine. The figure shows that the outer k-shells of the
transmission network (i.e., the 0.5-kshell defined as the union of the
k-shells with $k=1, 2, ..., \lceil{ 1/2 \, k_{\rm core}^{\rm max}
  \rceil}-1$, see SM sections~I-V) are disintegrated in the lockdown,
decreasing by 91\% with respect to their pre-quarantine size, in
tandem with the GCC. However, the inner k-core (i.e., the 0.5-kcore
defined as the k-core with $k=\lceil{1/2 \, k_{\rm core}^{\rm max}
  \rceil}$, see SM sections~I-V) persists in the lockdown. The figure shows
that the decrease of the 0.5-kcore is only 50\% compared to the 91\%
decrease of the 0.5-kshell; the former even increases slightly at the
end of April, following the same trend in mobility (see
Fig. \ref{gcc}b). This process is visually corroborated in the
evolution of the networks seen from Fig. \ref{networks}a to
\ref{networks}c where we observe the disappearance of the peripheral
k-shells and the persistence of the maximal k-core.  Indeed, the
unessential contacts in the peripheral k-shells may have been first
pruned during social distancing.



Using numerical simulations, we corroborate previous results
indicating that the infection can persist in these high k-cores of the
network while virus persistence in outer k-shells is less important
\cite{epidemic,kcore}. We use a SIR model on the transmission network
(Fig. \ref{sir-attack}a and SM Fig.~S14A) showing that
the maximal k-cores of the network sustain the spreading of the
disease more efficiently than the outer k-shells. Thus, the maximal
k-core components of the contact network are plausible drivers of
disease transmission. Apart from this structural explanation (i.e.,
k-core), epidemiological factors may also play a role in the
persistence of the disease,
such as a transition of the disease to vulnerable communities with
high demographic density, or with large inhabitants per household
where isolation is poorly fulfilled.


When we plot the geolocation of the contacts forming the maximal
k-core in the map of Cear\'a, we find that these contacts take place
in highly transited areas of the capital Fortaleza, such as hospitals,
business buildings, warehouses as well as large condominiums, see
Fig. \ref{networks}d.
These contacts generate superspreading k-core events that generalize
the conventional notion of superspreaders, which refer mainly to
individuals with large number of transmission contacts
\cite{may,superspreaders,superspreaders2}.  However, connections are
not everything \cite{barabasi-attack,shlomo}. K-core superspreaders
not only generate a large number of transmission contacts, but their
contacts are also highly connected people, and so forth.





\section{Optimized quarantine} 
The existence of k-cores in the
transmission network suggests that a more structured quarantine could
be deployed to either isolate or destroy those cores that help
maintain the spread of the virus. We perform an optimal percolation
analysis \cite{barabasi-attack,shlomo,ci} to find the minimal number
of people necessary to quarantine that will dismantle the transmission
network.   We compare different strategies to find the
  best among them to break the network by ranking the nodes based on
(1) the number of contacts (hub-removal)
\cite{barabasi-attack,shlomo,epidemic}, (2) the largest k-shells and
then by the degree inside the k-shells \cite{epidemic,kcore}, (3) the
collective influence algorithm for optimal percolation \cite{ci}, and
(4) betweenness centrality
\cite{freeman,friedkin91,barthelemy,dufresne} (we also try other
centralities, see SM sections~I-V).


Figure \ref{sir-attack}b shows the normalized size of the GCC versus
the fraction of removal nodes following different strategies, as well
as a random null model of removal in a typical network under lockdown
in April 28 (March 19 pre-lockdown results are plotted in SI
Fig.~S14B).  While the disease can persist in the
k-cores (Fig. \ref{sir-attack}a), quarantining people directly inside
the maximal k-core is not an optimal strategy. The reason is that
k-cores are populated by hyper-connected hubs that requiere many
removals to break the GCC \cite{dufresne} (around 7\%, see
Fig. \ref{sir-attack}b). For the same reason, removing directly the
hubs is not the optimal strategy either, since the hubs are within the
maximal k-core and not outside. A collective influence strategy
\cite{ci} improves over hub-removal since it takes into account how
hubs are spatially distributed, yet, it is far from optimal.  Clearly,
Fig. \ref{sir-attack}b shows that the best strategy is to quarantine
people by their betweenness centrality. By removing just the top
1.6-2\% of the high betweenness centrality people, the GCC is
disintegrated. This result is consistent with the particular structure
of the transmission networks seen in Fig. \ref{networks}b, c and
Fig. \ref{sir-attack}.

The betweenness centrality of a node is proportional to the number of
shortest paths in the network going through that node. Thus, given the
particular structure of the networks in Figs. \ref{networks}b, c, and
Fig. \ref{sir-attack}c, the high betweenness centrality nodes are the
bottlenecks of the network, i.e., loosely-connected bridges between
the largely-connected k-cores components. These connectors are the
'weak links', fundamental concept in sociology proposed by Granovetter
\cite{granovetter}, according to which, strong ties (i.e., contacts in
the k-cores) clump together forming clusters. A strategically located
weak tie between these densely 'knit clumps', then becomes the crucial
bridge that transmits the disease (or information \cite{granovetter})
between k-cores.
These weak links are people traveling among the different k-cores
components allowing the disease to escape the cores into the rest of
society.  These bridges are displayed in the network of
Fig. \ref{sir-attack}c as the yellow, blue and red nodes.  The removal
of these high betweenness centrality people disconnects the k-core
components of the network entirely, as shown in
Fig. \ref{sir-attack}d, halting the disease transmission from one core
to the other \cite{salathe,dufresne}.


An important finding is that quarantining the large superspreading
k-cores is neither optimal (as shown in Fig. \ref{sir-attack}b, green
curve) nor practical, since they are mainly comprised by chiefly
essential workers who need to remain operational
(Fig. \ref{networks}d). Thus, the best strategy, in conjunction with a
mass quarantine, is then to disconnect these k-cores from the rest of
the social network (Figs. \ref{sir-attack}c and \ref{sir-attack}d),
rather than quarantining the people inside the k-cores.  This can be
performed by quarantining the high betweenness centrality weak-links
that simultaneously preserve the operational k-cores.
However, individuals belonging to the maximal k-cores should be tested
at a higher frequency to promptly detect their infectiousness before
the symptoms start, to help control the spreading inside the k-cores.\\






\section{Summary}

Isolating the k-core structures by quarantining the
high betweenness centrality weak links in the transmission network
proves to be an effective way to dismantle the GCC of the
disease while keeping essential k-cores working. While destroying the
strong links and cores is a less manageable task to execute and
control, isolating the weak links between cores is a more feasible
task that will assure the dismantling of the GCC. In other words, if
one core is infected, the disease will be controlled within that core
and not extended to the rest of society.

  
As governments around the world are racing to roll out digital contact
tracing apps to curb the spread of coronavirus \cite{mit,DeVries,Singapore,Israel,Helsenorge,Noem,mit2,iceland}, our
modeling suggests possible
quarantine protocols that could become key in the second phase of
reopening economies across the world and, in particular, in developing
countries where resources are scarce.  Overall, our network-based
optimized protocol is reproducible in any setting and could become an
efficient solution to halt the critical progress of the COVID-19
pandemic worldwide drawing upon effective quarantines with minimal
disruptions.





\section*{Acknowledgments}
We are grateful to S. Alarc\'on-D\'iaz,
M. Sigman and I. Belausteguigoitia for discussions.
We also thank J. Maciel, G. Sousa, J. A. P. Barreto, and R. Sousa from
the Health Secretariat of Fortaleza for data curation.

\end{document}